\documentclass[12pt,preprint]{aastex}

\usepackage{epsfig}
\usepackage{graphicx}
\usepackage{lineno}

\def\us{\char`\_}

\slugcomment{To appear in ApJ}

\shorttitle{Multiwavelength Observations of the Previously Unidentified Blazar RX\,J0648.7+1516}
\shortauthors{Furniss et al.}

\begin{document}
\linenumbers

\title{Multiwavelength Observations of the Previously Unidentified Blazar RX\,J0648.7+1516}

\author{
E.~Aliu\altaffilmark{1},
T.~Aune\altaffilmark{2},
M.~Beilicke\altaffilmark{3},
W.~Benbow\altaffilmark{4},
M.~B\"ottcher\altaffilmark{5},
A.~Bouvier\altaffilmark{2},
S.~M.~Bradbury\altaffilmark{6},
J.~H.~Buckley\altaffilmark{3},
V.~Bugaev\altaffilmark{3},
A.~Cannon\altaffilmark{7},
A.~Cesarini\altaffilmark{8},
L.~Ciupik\altaffilmark{9},
M.~P.~Connolly\altaffilmark{8},
W.~Cui\altaffilmark{10},
G.~Decerprit\altaffilmark{11},
R.~Dickherber\altaffilmark{3},
C.~Duke\altaffilmark{12},
M.~Errando\altaffilmark{1},
A.~Falcone\altaffilmark{13},
Q.~Feng\altaffilmark{10},
G.~Finnegan\altaffilmark{14},
L.~Fortson\altaffilmark{15},
A.~Furniss\altaffilmark{2,*},
N.~Galante\altaffilmark{4},
D.~Gall\altaffilmark{16},
G.~H.~Gillanders\altaffilmark{8},
S.~Godambe\altaffilmark{14},
S.~Griffin\altaffilmark{17},
J.~Grube\altaffilmark{9},
G.~Gyuk\altaffilmark{9},
D.~Hanna\altaffilmark{17},
B.~Hivick\altaffilmark{5},
J.~Holder\altaffilmark{18},
H.~Huan\altaffilmark{19},
G.~Hughes\altaffilmark{11},
C.~M.~Hui\altaffilmark{14},
T.~B.~Humensky\altaffilmark{19},
P.~Kaaret\altaffilmark{16},
N.~Karlsson\altaffilmark{15},
M.~Kertzman\altaffilmark{20},
D.~Kieda\altaffilmark{14},
H.~Krawczynski\altaffilmark{3},
F.~Krennrich\altaffilmark{21},
G.~Maier\altaffilmark{11},
P.~Majumdar\altaffilmark{22},
S.~McArthur\altaffilmark{3},
A.~McCann\altaffilmark{17},
P.~Moriarty\altaffilmark{23},
R.~Mukherjee\altaffilmark{1},
T.~Nelson\altaffilmark{30},
R.~A.~Ong\altaffilmark{22},
M.~Orr\altaffilmark{21},
A.~N.~Otte\altaffilmark{2},
N.~Park\altaffilmark{19},
J.~S.~Perkins\altaffilmark{24,25},
A.~Pichel\altaffilmark{26},
M.~Pohl\altaffilmark{27,11},
H.~Prokoph\altaffilmark{11},
J.~Quinn\altaffilmark{7},
K.~Ragan\altaffilmark{17},
L.~C.~Reyes\altaffilmark{19},
P.~T.~Reynolds\altaffilmark{28},
E.~Roache\altaffilmark{4},
H.~J.~Rose\altaffilmark{6},
J.~Ruppel\altaffilmark{27,11},
D.~B.~Saxon\altaffilmark{18},
G.~H.~Sembroski\altaffilmark{10},
C.~Skole\altaffilmark{11},
A.~W.~Smith\altaffilmark{29},
D.~Staszak\altaffilmark{17},
G.~Te\v{s}i\'{c}\altaffilmark{17},
M.~Theiling\altaffilmark{4},
S.~Thibadeau\altaffilmark{3},
K.~Tsurusaki\altaffilmark{16},
J.~Tyler\altaffilmark{17},
A.~Varlotta\altaffilmark{10},
V.~V.~Vassiliev\altaffilmark{22},
S.~P.~Wakely\altaffilmark{19},
T.~C.~Weekes\altaffilmark{4},
A.~Weinstein\altaffilmark{21},
D.~A.~Williams\altaffilmark{2},
B.~Zitzer\altaffilmark{10}
(The VERITAS Collaboration)\\
S.~Ciprini\altaffilmark{33},
M.~Fumagalli\altaffilmark{31,*},
K.~Kaplan\altaffilmark{4},
D.~Paneque\altaffilmark{34,*},
J.~X.~Prochaska\altaffilmark{32}
}

\altaffiltext{*}{Corresponding authors: A.~Furniss: afurniss@ucsc.edu, D.~Paneque: 
dpaneque@mppmu.mpg.de, M.~Fumagalli: miki@ucolick.org}
\altaffiltext{1}{Department of Physics and Astronomy, Barnard College, Columbia University, NY 10027, USA}
\altaffiltext{2}{Santa Cruz Institute for Particle Physics and Department of Physics, University of California, Santa Cruz, CA 95064, USA}
\altaffiltext{3}{Department of Physics, Washington University, St. Louis, MO 63130, USA}
\altaffiltext{4}{Fred Lawrence Whipple Observatory, Harvard-Smithsonian Center for Astrophysics, Amado, AZ 85645, USA}
\altaffiltext{5}{Astrophysical Institute, Department of Physics and Astronomy, Ohio University, Athens, OH 45701, USA}
\altaffiltext{6}{School of Physics and Astronomy, University of Leeds, Leeds, LS2 9JT, UK}
\altaffiltext{7}{School of Physics, University College Dublin, Belfield, Dublin 4, Ireland}
\altaffiltext{8}{School of Physics, National University of Ireland Galway, University Road, Galway, Ireland}
\altaffiltext{9}{Astronomy Department, Adler Planetarium and Astronomy Museum, Chicago, IL 60605, USA}
\altaffiltext{10}{Department of Physics, Purdue University, West Lafayette, IN 47907, USA }
\altaffiltext{11}{DESY, Platanenallee 6, 15738 Zeuthen, Germany}
\altaffiltext{12}{Department of Physics, Grinnell College, Grinnell, IA 50112-1690, USA}
\altaffiltext{13}{Department of Astronomy and Astrophysics, 525 Davey Lab, Pennsylvania State University, University Park, PA 16802, USA}
\altaffiltext{14}{Department of Physics and Astronomy, University of Utah, Salt Lake City, UT 84112, USA}
\altaffiltext{15}{School of Physics and Astronomy, University of Minnesota, Minneapolis, MN 55455, USA}
\altaffiltext{16}{Department of Physics and Astronomy, University of Iowa, Van Allen Hall, Iowa City, IA 52242, USA}
\altaffiltext{17}{Physics Department, McGill University, Montreal, QC H3A 2T8, Canada}
\altaffiltext{18}{Department of Physics and Astronomy and the Bartol Research Institute, University of Delaware, Newark, DE 19716, USA}
\altaffiltext{19}{Enrico Fermi Institute, University of Chicago, Chicago, IL 60637, USA}
\altaffiltext{20}{Department of Physics and Astronomy, DePauw University, Greencastle, IN 46135-0037, USA}
\altaffiltext{21}{Department of Physics and Astronomy, Iowa State University, Ames, IA 50011, USA}
\altaffiltext{22}{Department of Physics and Astronomy, University of California, Los Angeles, CA 90095, USA}
\altaffiltext{23}{Department of Life and Physical Sciences, Galway-Mayo Institute of Technology, Dublin Road, Galway, Ireland}
\altaffiltext{24}{CRESST and Astroparticle Physics Laboratory NASA/GSFC, Greenbelt, MD 20771, USA.}
\altaffiltext{25}{University of Maryland, Baltimore County, 1000 Hilltop Circle, Baltimore, MD 21250, USA.}
\altaffiltext{26}{Instituto de Astronomia y Fisica del Espacio, Casilla de Correo 67 - Sucursal 28, (C1428ZAA) Ciudad Aut—noma de Buenos Aires, Argentina}
\altaffiltext{27}{Institut f\"ur Physik und Astronomie, Universit\"at Potsdam, 14476 Potsdam-Golm,Germany}
\altaffiltext{28}{Department of Applied Physics and Instrumentation, Cork Institute of Technology, Bishopstown, Cork, Ireland}
\altaffiltext{29}{Argonne National Laboratory, 9700 S. Cass Avenue, Argonne, IL 60439, USA}

\altaffiltext{30}{School of Physics and Astronomy, University of Minnesota, 116 Church St. SE, Minneapolis, MN 55455, USA}
\altaffiltext{31}{Department of Astronomy and Astrophysics, University of California, 1156 High Street, Santa Cruz, CA 95064}
\altaffiltext{32}{Department of Astronomy and Astrophysics, UCO/Lick Observatory, University of California, 1156 High Street, Santa Cruz, CA 95064}
\altaffiltext{33}{Dipartimento di Fisica, Universit\`a degli Studi di Perugia, I-06123 Perugia, Italy}
\altaffiltext{34}{Max-Planck-Institut f\"ur Physik, D-80805 M\"unchen, Germany}

\begin{abstract}
We report on the VERITAS discovery of very-high-energy (VHE) gamma-ray emission above 200 GeV from the high-frequency-peaked BL Lac object RX\,J0648.7+1516 (GB\,J0648+1516), associated with 1FGL\,J0648.8+1516.  The photon spectrum above 200 GeV is fit by a power law $dN/dE = F_0 (E/E_0)^{-\Gamma}$ with a photon index $\Gamma$ of $4.4 \pm 0.8_{stat} \pm 0.3_{syst}$ and a flux normalization $F_0$ of $(2.3 \pm 0.5_{stat} \pm 1.2_{sys}) \times 10^{-11} $ TeV$^{-1} $cm$^{-2} $s$^{-1}$ with $E_0 =$ 300 GeV.  No VHE variability is detected during VERITAS observations of RX\,J0648.7+1516 between 2010 March 4 and April 15.  Following the VHE discovery, the optical identification and spectroscopic redshift were obtained using the Shane 3--m Telescope at the Lick Observatory, showing the unidentified object to be a BL Lac type with a redshift of z $= 0.179$.   Broadband multiwavelength observations contemporaneous with the VERITAS exposure period can be used to sub-classify the blazar as a high-frequency-peaked BL Lac (HBL) object, including data from the MDM observatory, \textit{Swift}-UVOT and XRT, and continuous monitoring at photon energies above 1 GeV from the \textit{Fermi} Large Area Telescope (LAT).  We find that in the absence of undetected, high-energy rapid variability, the one-zone synchrotron self-Compton model (SSC) overproduces the high-energy gamma-ray emission measured by the \textit{Fermi}-LAT over 2.3 years.  The SED can be parameterized satisfactorily with an external-Compton or lepto-hadronic model, which have two and six additional free parameters, respectively, compared to the one-zone SSC model.
\end{abstract}


\section{Introduction}

1FGL\,J0648.8+1516 was detected by \textit{Fermi}-LAT in the first 11 months of operation at greater than 10 standard deviations, $\sigma$ \citep{1FGL}. This source was flagged as a very-high-energy (VHE; E$>$100 GeV) emitting candidate by the \textit{Fermi}-LAT collaboration by searching for $\ge$30\,GeV photons.  This information triggered the VERITAS observations reported here. 1FGL\,J0648.8+1516 is found to be associated with RX\,J0648.7+1516, which was first discovered by ROSAT \citep{brinkmann}.  A radio counterpart was identified in the NRAO Green Bank survey \citep{becker}.   Two subsequent attempts to identify an optical counterpart were unsuccessful \citep{motch,haakonsen}.

At 6$^{\circ}$ off the Galactic plane and without optical spectroscopy, the nature of this object remained unknown until optical spectroscopy was obtained in response to the VERITAS detection.  These observations allow the active galactic nucleus (AGN) to be classified as a BL Lac, a type of AGN that has a jet co-aligned closely with the Earth's line of sight and displays weak emission lines.  These AGN are characterized by non-thermal, double-peaked broadband spectral energy distributions (SED).  Based on the radio and X-ray flux, the BL Lac can further be classified as a high-frequency-peaked BL Lac (HBL) \citep{padovani}, or if classified by the location of its low-energy peak, a high-synchrotron-peaked BL Lac (HSP) \citep{abdoSED}.

\section{Observations and Analysis} 
\subsection{VERITAS}
VERITAS comprises four imaging atmospheric Cherenkov telescopes and is sensitive to gamma-rays between $\sim$100 GeV and $\sim$30 TeV \citep{weekes2002,holder2006}. The VERITAS observations of RX\,J0648.7+1516 were completed between 2010 March 4 and April 15 (MJD 55259-55301), resulting in 19.3 hours of quality-selected live time.  These observations were taken at 0.5$^{\circ}$ offset in each of four directions to enable simultaneous background estimation using the reflected-region method \citep{fomin}. 

The VERITAS events are parameterized by the principal moments of the elliptical shower images, allowing cosmic-ray background rejection through a set of selection criteria (cuts) which have been optimized \textit{a priori} on a simulated, soft-spectrum (photon index 4.0) source with a VHE flux 6.6\% of that observed from the Crab Nebula.   
The cuts discard images with fewer than $\sim$50 photoelectrons.  Events with at least two telescope images remaining are then cosmic-ray discriminated based on the mean-scaled-width (MSW) and the mean-scaled-length (MSL) parameters.  Events with MSW $<$ 1.1, MSL $<$ 1.4, a height of maximum Cherenkov emission $>$ 8 km and an angular distance to the reconstructed source position in the camera ($\theta$) of less than 0.14 degrees are kept as gamma-ray candidate events.  The results are reproduced in two independent analysis packages \citep{cogan, daniel}. 
After background rejection, 2711 events remain in the source region, with 16722 events remaining in the background regions (larger by a factor of 6.89).   The 283 excess events result in a significance of 5.2$\sigma$, calculated using Equation 17 from \cite{lima}.

A differential power law $dN/dE=F_o (E/300$ GeV$)^{-\Gamma}$ is fit to the VERITAS data from 200 to 650\,GeV, shown in the top panel of Figure 1.  The fit ($\chi^2 = 0.90$ with 3 degrees of freedom (DOF), probability of 0.83) results in a flux normalization of $F_o = (2.3 \pm 0.5_{stat} \pm 1.2_{syst}) \times 10^{-11}$ photons cm$^{-2}$ s$^{-1}$ TeV$^{-1}$ and an index of $\Gamma = 4.4 \pm 0.8_{stat} \pm 0.3_{syst}$, corresponding to 3.3\% of the Crab Nebula flux above 200 GeV.

The angular distribution of the excess events is consistent with a point source now designated VER\,J0648+152, located at 102.19$^{\circ}$ $\pm$ 0.11$^{\circ}_{stat}$ RA and 15.27$^{\circ}$ $\pm$ 0.12$^{\circ}_{stat}$ Dec (J2000).  The systematic pointing uncertainty of VERITAS is less than 25$\arcsec$ (7$\times10^{-3}$ degrees).  This position is consistent with the radio position of RX\,J0648.7+1516 \citep{becker}.  A nightly-binned VHE light curve is fit with a constant and shows a $\chi^2$ null hypothesis probability of 0.39, showing no significant variability during the observation.

\subsection{\textit{Fermi}-LAT}
The \textit{Fermi}-LAT is a pair-conversion telescope sensitive to photons between 20 MeV and several hundred GeV \citep{Atwood2009, abdo2009}. The data used in this paper encompass the time interval 2008 Aug 5 through 2010 Nov 17 (MJD 54683-55517), and were analyzed with the LAT \texttt{ScienceTools} software package version \texttt{v9r15p6}, which is available from the Fermi Science Support Center (FSSC). Only events from the ``diffuse" class with energy above 1 GeV within a 5$^{\circ}$ radius of RX\,J0648.7+1516 and with a zenith angle $< 105^{\circ}$ were used. The background was parameterized with the files gll\_iem\_v02.fit and isotropic\_iem\_v02.txt \footnote{The files are available at \url{http: //fermi.gsfc.nasa.gov/ssc/data/access/lat/ BackgroundModels.html}}. The normalizations of the components were allowed to vary freely during the spectral point fitting, which was performed with the unbinned likelihood method and using the instrument response function \texttt{P6\_V3\_DIFFUSE}.

The spectral fits using energies above 1 GeV are less sensitive to possible contamination from unaccounted (transient) neighboring sources, and hence have smaller systematic errors, at the expense of slightly reducing the number of source photons. Additionally, there is no significant signal from RX\,J0648.7+1516 below 1 GeV.  The analysis of 2.3 years between 2008 Aug 5 and 2010 Nov 17 (MJD 54683--55517) of \textit{Fermi}-LAT events with energy between 0.3--1 GeV (fixing the spectral index to 1.89) yields a test statistic (TS) of 9, corresponding to $\sim 3 \sigma$ \footnote{See \cite{mattox} for TS definition.}. In addition to the background, the emission model includes two nearby sources from the 1FGL catalog: the pulsars PSR\,J0659+1414 and PSR\,J0633+1746.  The spectra from the pulsars are parameterized with power-law functions with exponential cutoffs, and the values are fixed to the values found from 18 months of data. The
spectral fluxes are determined using an unbinned maximum
likelihood method.  The flux systematic uncertainty is estimated as $5\%$ at $560$\,MeV
and $20\%$ at $10$\,GeV and above.\footnote{See
\texttt{http://fermi.gsfc.nasa.gov/ssc/data/analysis/LAT\us
caveats.html}}   

The results from the \textit{Fermi}-LAT spectral analysis are shown in the bottom panel of Figure 1. There is no variability detected in four time bins evenly spread over the 2.3 years of data.  The dataset corresponding in time to the VERITAS observations between between 2010 March 4 and April 15 (i.e. MJD 55259$-$55301) does not show any significant signal and thus we report 2$\sigma$ upper limits that were computed using the Bayesian method \citep{helene}, where the likelihood is integrated from zero up to the flux that encompasses 95\% of the posterior probability.  When using the data accumulated over the expanded full 2.3 years of data, we find that 1FGL\,J0648.8+1516 is significantly detected above 1\,GeV with a TS of 307. The spectrum is fit using a single power-law function with photon flux $F_{>1\,GeV} = (1.8 \pm  0.2_{stat})  \times 10^{-9}$ photons cm$^{-2}  $s$^{-1}$  and hard differential photon spectral index  $\Gamma_{LAT} = 1.89 \pm  0.10_{stat}$. The analysis is also performed on five energy ranges equally spaced on a log scale with the photon index fixed to 1.89 and only fitting the normalization. The source is detected significantly (TS$>$25) in each energy bin except for the highest energy (100-300 GeV), for which a 95\% confidence level upper limit is calculated.

\subsection{\textit{Swift}-XRT}
The \textit{Swift}-XRT \citep{gehrels,burrows05} data are analyzed with HEASOFT\,6.9 and XSPEC version 12.6.0. Observations were taken in photon counting mode with an average count rate of $\sim 0.3$ counts per second and did not suffer from pile-up.  Six target-of-opportunity observations summing to 10.5 ks were collected on six different days between 2010 March 18 and April 18 (MJD 55273 and 55304), inclusive.  These observations were combined with a response file created from summing each observation's exposure file using \textit{ximage}.  The photons are grouped by energy to require a minimum of 30 counts per bin, and fit with an absorbed power law between 0.3 and 10 keV, allowing the neutral hydrogen (HI) column density to vary.  A HI column density of $1.94 \pm 0.14 \times 10^{21} $cm$^{-2}$ is found, only slightly higher than the $1.56 \times 10^{21} $cm$^{-2}$ quoted in \cite{kalberla}.  The combined X-ray energy spectrum is extracted with a fit ($\chi^2 = 114$ for 88 DOF, null hypothesis probability of 3.2$\times10^{-2}$) with a photon index of $2.51 \pm 0.06$ and an integral flux between 0.3 and 10 keV of $(1.24 \pm 0.03_{\textrm{stat}}) \times 10^{-11}$ ergs cm$^{-2}$ s$^{-1}$.  This corresponds to a 0.3 to 10 keV rest frame luminosity of $1.1 \times 10^{45}$ ergs s$^{-1}$.  The deabsorbed spectrum is used to constrain modeling.

\subsection{\textit{Swift}-UVOT}
The \textit{Swift}-XRT observations were supplemented with UVOT exposures taken in the U, UVM2, and UVW2 bands (centered at $8.56 \times 10^{14}$\,Hz, $1.34 \times 10^{15}$\,Hz, and $1.48 \times 10^{15}$\,Hz, respectively; \cite{poole}).  The UVOT photometry is performed using the HEASOFT program \textit{uvotsource}.  The circular source region has a $5\arcsec$ radius and the
background regions consist of several circles with radii between $10-15 \arcsec$ of nearby empty sky.
The results are reddening corrected using R(V)=3.32 and E(B-V)=0.14 \citep{schlegel}.  The Galactic extinction coefficients were applied according to \cite{fitzpatrick}, with the largest source of error resulting from deredenning.  A summary of the UVOT analysis results is given in Table 1.

\subsection{Optical MDM}
The region around RX\,J0648.7+1516 was observed in the optical B, V, and R bands with the 1.3-m McGraw-Hill Telescope of the MDM Observatory
on four nights during 2010 April 1--5 (MJD 55287-55291). Exposure times ranged from 90 sec (R-band) to 120 sec (B-band). Each night, five sequences of exposures in B, V, and R were taken.  The raw data were bias subtracted and flat-field corrected using standard routines in IRAF\footnote{\tt http://www.noao.edu/credit.html}.  Aperture photometry is performed using the IRAF package DAOPHOT on the object as well as five comparison stars in the same field of view. Calibrated magnitudes of the comparison stars are taken from the NOMAD catalog\footnote{\tt http://www.nofs.navy.mil/nomad.html},
and the magnitudes of the object are determined using comparative photometry methods.  For the construction of the SED points, the magnitudes are extinction corrected based on the \cite{schlegel} dust map with values taken from NASA Extragalactic Database (NED)\footnote{\tt http://nedwww.ipac.caltech.edu/} : $A_B = 0.618$, $A_V = 0.475$, and $A_R = 0.383$.   These data (summarized in Table 1) are used to constrain the modeling shown in this work, although the same conclusions result with the UVOT points as model constraint.

\section{Spectroscopic Redshift Measurements}
Two spectra were obtained during the nights of UT 2010 March 18 and 2010 November 6 (MJD 55245 and 55506, respectively)
with the KAST double spectrograph on the Shane 3-m Telescope at UCO/Lick Observatory. 
During the first night, the instrument was configured with a 600/5000 grating and $1.5\arcsec$ long slit, covering $4300-7100$ \AA.  A single 1800 second exposure was acquired. During the night of November 6, another 1800 second exposure was acquired with a 600/4310 grism,  D55 dichroic, a $600/7500$ grating and $2\arcsec$ long slit, covering the interval $3500-8200$ \AA.  The data were reduced with the LowRedux pipeline\footnote{\tt http://www.ucolick.org/$\sim$xavier/LowRedux/index.html} and flux calibrated using a spectro-photometric star. The flux calibration is uncertain due to non-photometric conditions.
Inspection of the March spectrum reveals Ca H+K absorption lines at redshift $z = 0.179$. This redshift is confirmed in the second spectrum at higher signal-to-noise (S/N) (S/N $\sim 20$ in the blue and S/N $\sim 50$ in the red) where Ca H+K, G band, Mg\,I $\lambda \lambda \lambda~5168,5174,5184$ and Na\,I $\lambda \lambda \lambda~5891,5894,5897$ absorption lines with equivalent width $< 5$ \AA ~are detected (see Figure \ref{fig2} and Table 2 for details).  No Ca H+K break is observed. These spectral features provide evidence for an early-type nature of the blazar host galaxy and allow for BL Lac classification, following \cite{marcha} and \cite{healey}.

\section{Broadband SED Modeling}
The contemporaneous multiwavelength data are matched with archival radio data from NED and are shown in Figure 3.  Since the radio data are not contemporaneous they are shown only for reference. The synchrotron peak appears at a frequency greater than $10^{16}$ Hz, representing the first subclassification of RX\,J0648.7+1516, specifically as an HBL. These data are
used to test steady-state leptonic and lepto-hadronic jet models for
the broadband blazar emission.  The absorption of VHE gamma rays by the extragalactic background light (EBL) is accounted for through application of the \cite{gilmore09} EBL model; the model of \cite{finke} provides comparable results.

Leptonic models for blazar emission attribute the higher-energy
peak in the SED to the inverse-Compton scattering of lower-energy photons off a population of non-thermal, relativistic
electrons.  These same electrons are responsible for the lower-energy
synchrotron emission making up the first peak.  The target photon
field involved in the Compton upscattering can either be the synchrotron
photons themselves, as in synchrotron self-Compton (SSC) models, or a photon field external to the jet in
the case of external Compton (EC) models.  

We use the equilibrium SSC model of
\cite{bc02}, as described in \cite{acciari09}.  In this model,
the emission originates from a spherical blob of relativistic
electrons with radius $R$.  This blob is moving down the jet with a
Lorentz factor $\Gamma$, corresponding to a jet speed of
$\beta_{\Gamma} c$.  The jet is oriented such that the angle
with respect to the line of sight is $\theta_{\rm obs}$, which results
in a Doppler boosting with Doppler factor $D = (\Gamma [1-\beta_{\Gamma}
\cos\theta_{\rm obs}])^{-1}$.  In order to minimize the number of
free parameters, the modeling is completed with $\theta_{obs} = 1/\Gamma$,
for which $\Gamma = D$.

Within the model, electrons are injected with a power-law distribution at a rate $Q(\gamma) = Q_0
\gamma^{-q}$ between the low- and high-energy cut-offs,
$\gamma_{1,2}$. The electron spectral index of $q = 4.8$ required for the models applied in this work might be the result of
acceleration in an oblique shock. While standard shock acceleration in relativistic,
parallel shocks is known to produce a canonical spectral index of $\sim$2.2, oblique
magnetic-field configurations reduce the acceleration efficiency and lead to much
steeper spectral indices \citep{meli, sironi}. The radiation mechanisms considered lead to
equilibrium between the particle injection, radiative cooling and
particle escape.  The particle escape is characterized with an
efficiency factor $\eta$, such that the escape timescale $t_{\rm
esc} = \eta \, R/c$, with $\eta=100$ for this work.  This results in a particle distribution streaming along the jet with a power $L_e$.  Synchrotron emission
results from the presence of a tangled magnetic field $B$, with a
Poynting flux luminosity of $L_B$.  The parameters $L_e$ and
$L_B$ allow the calculation of the equipartition parameter
$\epsilon_{Be} \equiv L_B/L_e$. 

The top panel in Figure 3 shows the SSC model for RX\,J0648.7+1516, with parameters summarized in Table 3.  
The model is marginally in agreement with the data only through use of parameters well below equipartition. The \textit{Fermi}-LAT contemporaneous 95\% confidence level upper limits in the energy ranges 1-3 GeV and
3-10 GeV are just above and below the one-zone SSC model
predictions. Additionally, these SSC model predictions are above the 2.3
year \textit{Fermi}-LAT spectrum by more than a factor of 2, although this
spectrum is not contemporaneous with the other data.  Variation of the model parameters within physically reasonable values does not provide better agreement between model and data.  Generally, HBLs are well characterized by one-zone SSC models and hence these observations might suggest the existence of one or more additional emission mechanisms that contribute to the higher-energy peak. 

An external-Compton model is also used to describe the data.  
The EC model is a leptonic one-zone jet model with two additional parameters beyond the SSC parameters, the thermal blackbody temperature $T_{EC}$ and radiation energy density $u_{EC}$ of the external photon field, which is assumed to be isotropic and stationary in the blazar rest frame.  The EC model provides a better representation of the SED, as can be seen in the middle panel of Figure 3, with the parameters listed in Table 3.  

A lepto-hadronic model is also applied to the data.  Within this model, ultrarelativistic protons are
the main source of the high-energy emission through proton synchrotron
radiation and pion production.  The resulting spectra of the pion
decay products are
evaluated with the templates of \cite{ka08}.  Additionally, a
semi-analytical description is used to account for electromagnetic
cascades initiated by the internal $\gamma\gamma$ absorption of
multi-TeV photons by both the $\pi^0$ decay photons and the synchrotron emission of ultrarelativistic leptons, as explained
in \cite{boettcher10}.  Similar to the particle populations in the leptonic models described above,
this lepto-hadronic model assumes a power-law distribution of
relativistic protons, $n(\gamma) \propto \gamma^{-q}$ between a low-
and high-energy cut-off, $E_{\rm p}^{\rm min,max}$.  This population
of relativistic protons is propagating along the blazar jet and has a
total kinetic luminosity of $L_p$.  The lepto-hadronic modeling results are above $\epsilon_{Bp}$ equipartition and are shown in the bottom panel of Figure 3 with parameters (including energy partition fractions $\epsilon_{Bp} \equiv L_B/L_p$ and $\epsilon_{ep} \equiv L_e/L_p$) summarized in Table 3. 

In conclusion, multiwavelength followup of the VERITAS detection of 1FGL\,J0648.7+1516 has solidified its association with RX\,J0648.7+1516, which is identified as a BL Lac object of the HBL subclass.  Other contemporaneous SEDs of VHE-detected HBLs can be well described by one-zone SSC models close to equipartition, while for RX\,J0648.7+1516 this model provides a poor representation with parameters below equipartition.  The addition of an external photon field for Compton up-scattering in the leptonic paradigm provides a better representation of the gamma-ray (\textit{Fermi} and VERITAS) data.  Alternatively, a lepto-hadronic model is successful in characterizing the higher-energy peak of the SED with synchrotron emission from protons.  Both of these latter models require super-equipartition conditions.

\acknowledgments

The authors of the paper thank the ApJ referee for the well organized and constructive comments that helped to improve
the quality and clarity of this publication.

VERITAS is supported by the US Department of Energy, US National Science Foundation and Smithsonian Institution, by NSERC in Canada, by Science Foundation Ireland (SFI 10/RFP/AST2748), and STFC in the UK.  We acknowledge the excellent work of the technical support staff at the FLWO and at the collaborating institutions.  This work was also supported by NASA grants from the Swift (NNX10AF89G) and Fermi (NNX09AU18G) Guest Investigator programs.

The \textit{Fermi} LAT Collaboration acknowledges generous support
from a number of agencies and institutes that have supported the
development and the operation of the LAT as well as scientific data analysis.
These include the National Aeronautics and Space Administration and the
Department of Energy in the United States, the Commissariat \`a l'Energie Atomique
and the Centre National de la Recherche Scientifique / Institut National de Physique
Nucl\'eaire et de Physique des Particules in France, the Agenzia Spaziale Italiana
and the Istituto Nazionale di Fisica Nucleare in Italy, the Ministry of Education,
Culture, Sports, Science and Technology (MEXT), High Energy Accelerator Research
Organization (KEK) and Japan Aerospace Exploration Agency (JAXA) in Japan, and
the K.~A.~Wallenberg Foundation, the Swedish Research Council and the
Swedish National Space Board in Sweden.

Additional support for science analysis during the operations phase is
acknowledged from the Istituto Nazionale di Astrofisica in Italy and the Centre National d'\'Etudes Spatiales in France.

J.X.P. acknowledges
funding through an NSF CAREER grant (AST--0548180).

{\it Facilities:} \facility{VERITAS}, \facility{Fermi}, \facility{Swift}, \facility{Lick}, \facility{MDM}.

\clearpage

\begin{figure}
\plotone{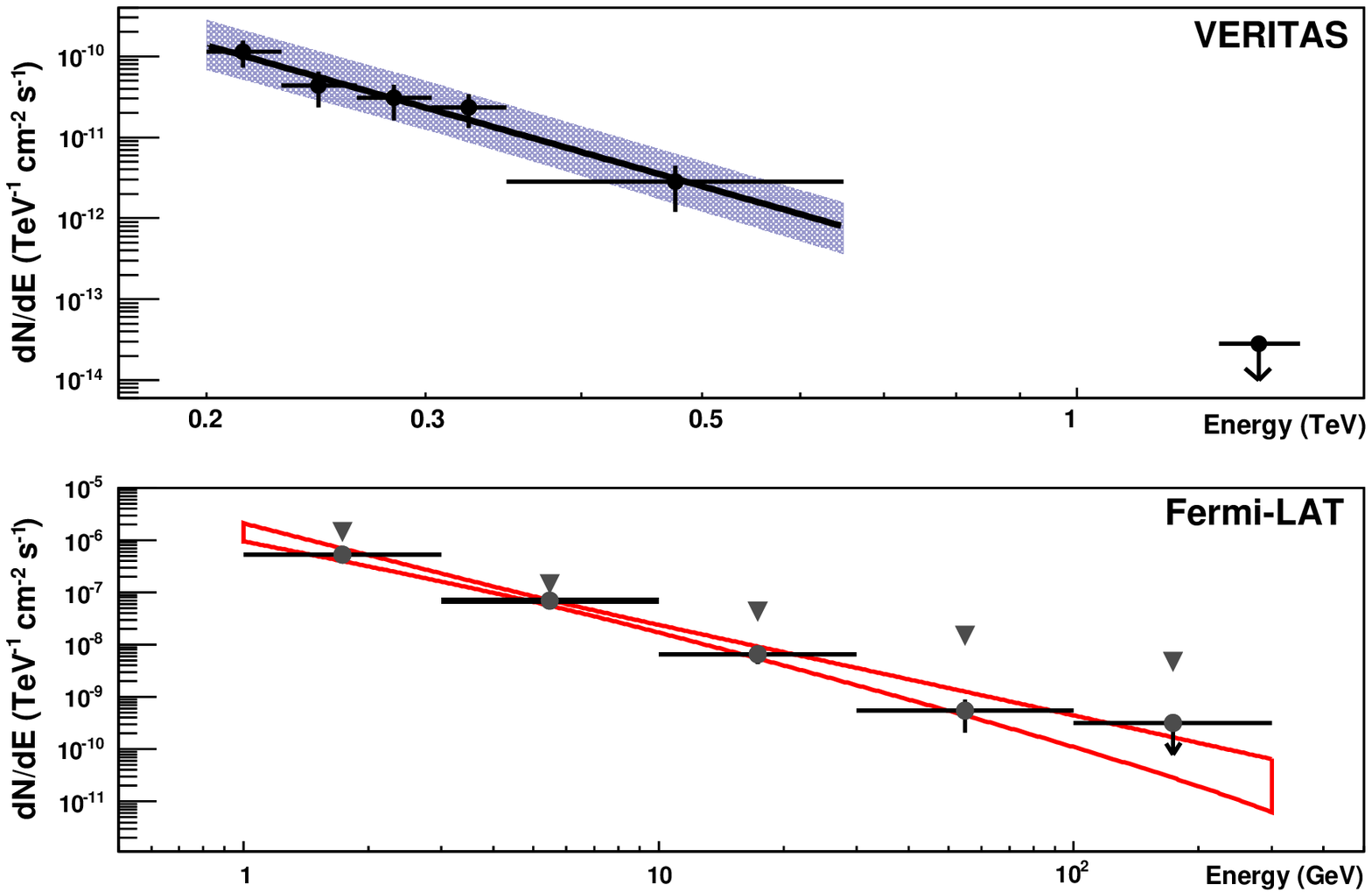}
\caption{Top: The differential photon spectrum of RX\,J0648.7+1516 between 200 and 650 GeV measured by VERITAS between 2010 4 March and 15 April (MJD 55259--55301).  The solid line shows a power-law fit to the measured flux derived with four equally log-spaced bins and a final bin boundary at 650 GeV, above which there are few on-source photons.  A 99\% confidence upper limit evaluated between 650 GeV and 5 TeV assuming a photon index of 4.4 is also shown.  The shaded region shows the systematic uncertainty of the fit, which is dominated by 20\% uncertainty on the energy scale. Bottom:  The differential photon spectrum of RX\,J0648.7+1516 as measured by \textit{Fermi}-LAT over 2.3 years between 2008 Aug 5 and 2010 Nov 17 (MJD 54683--55517, grey circles) with the highest energy bin containing a 95\%  confidence upper limit.  \textit{Fermi}-LAT upper limits from the VERITAS observation period are also shown (MJD 55259--55301, grey triangles). \label{fig1}}
\end{figure}

\clearpage
\begin{figure}
\includegraphics[angle=90, scale=0.6]{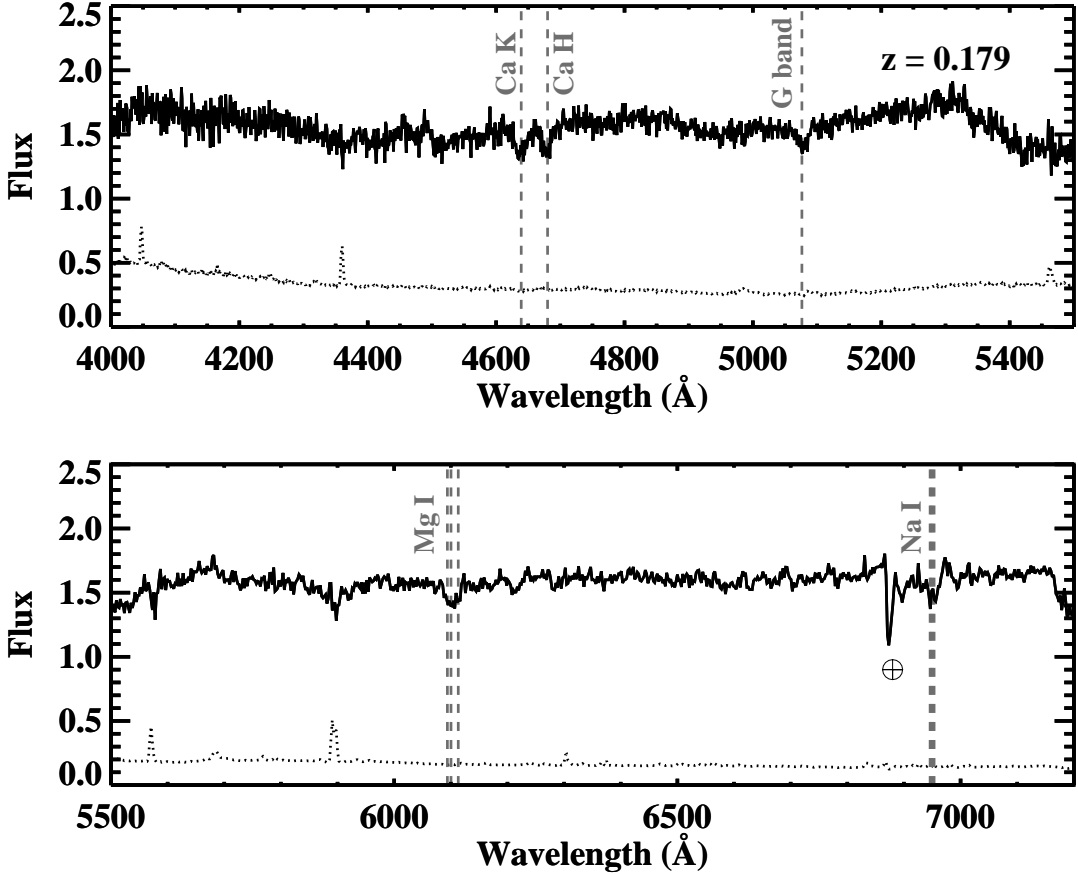}
\caption{Spectrum of RX\,J0648.7+1516 showing the Ca H+K, G-band, Na\,I and Mg\,I spectral features indicating a redshift of $z = 0.179$. Since the G-band arises in stellar atmospheres, we interpret this as the redshift for the host galaxy and not an intervening absorber. The blazar was
observed at Lick Observatory using the 3$-$m Shane Telescope on 6 November 2010. \label{fig2}
}
\end{figure}
\clearpage

\begin{figure}
\begin{center}
\epsscale{1.1}
\includegraphics[scale=0.65]{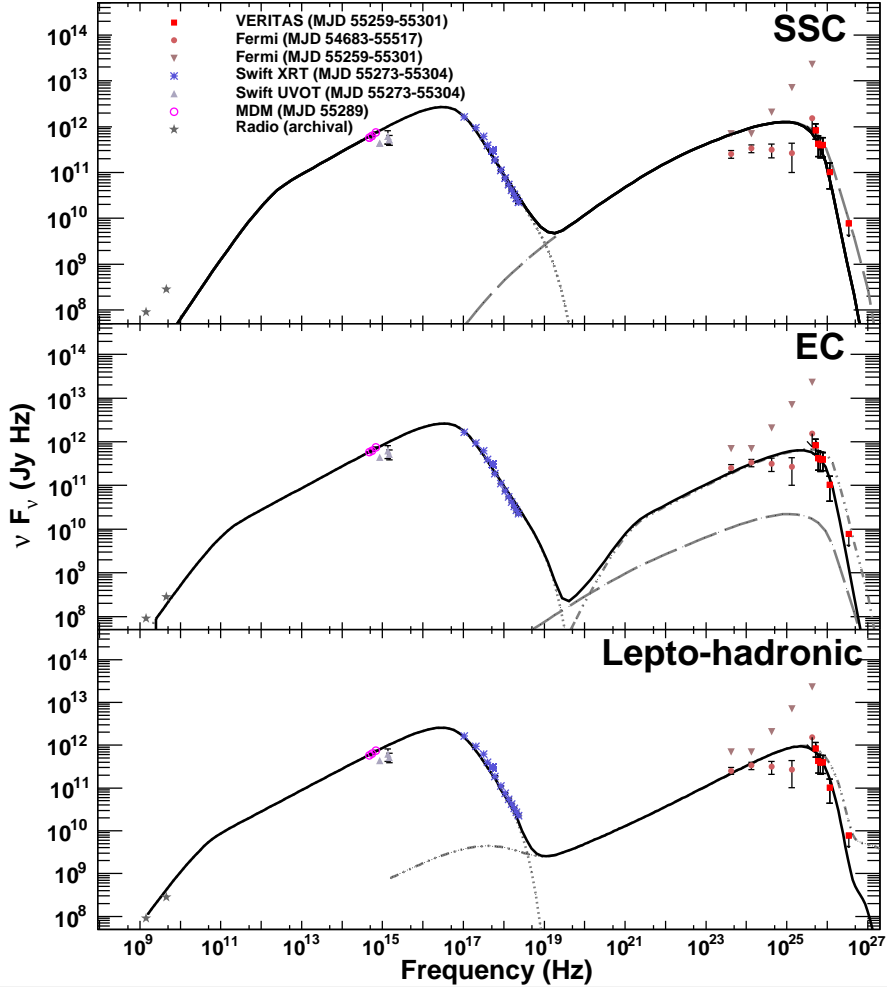}
\caption{The SED models applied to the contemporaneous multiwavelength data of RX\,J0648.7+1516. \textit{Fermi}-LAT data points are shown for 2.3 years of data along with upper limits extracted from data limited to the VERITAS observation period.  The models shown here are constrained by the MDM points; modeling constrained by the UVOT data produces similar results.  The top panel shows the synchrotron emission (dotted line), the self-Compton emission (dashed) and the EBL-corrected \citep{gilmore09} total one-zone SSC model (solid).  The middle panel shows the synchrotron emission (dotted line), the self-Compton emission (dashed line), the external-Compton (dash-dotted line) and the EBL-corrected total EC model (solid).  The bottom panel shows the electron (and positron) synchrotron emission (dotted line), the proton synchrotron emission (dash-dotted) and the EBL-corrected total lepto-hadronic model (solid).\label{fig3}}
\end{center}
\end{figure}
\clearpage

\begin{deluxetable}{cccc}
\tabletypesize{\scriptsize}
\tablecaption{Analysis summary of the optical MDM (B, V, R) and \textit{Swift}-UVOT (U, UVM2, UVW2) data. \label{tbl-1}}
\tablewidth{0pt}
\tablehead{
 \colhead{Band} &\colhead{Date} & \colhead{$\nu$F$_{\nu}$} & \colhead{$\nu$F$_{\nu}$ Error} \\
 \colhead{} &\colhead{(MJD)} & \colhead{(Jy Hz)} & \colhead{(Jy Hz)}\\}
\startdata
B & 55287 & 7.47$\times10^{11}$ & 3.4$\times10^{10}$\\
B & 55289 & 7.64$\times10^{11}$ & 3.8$\times10^{10}$\\
B & 55290 & 5.75$\times10^{11}$ & 2.7$\times10^{10}$\\
B & 55291 & 7.59$\times10^{11}$ & 3.4$\times10^{10}$\\
V & 55287 & 5.77$\times10^{11}$ & 3.5$\times10^{10}$\\
V & 55289 & 5.74$\times10^{11}$ & 3.7$\times10^{10}$\\
V & 55290 & 2.92$\times10^{11}$ & 1.6$\times10^{10}$\\
V & 55291 & 6.00$\times10^{11}$ & 3.6$\times10^{10}$\\
R & 55287 & 5.99$\times10^{11}$ & 4.2$\times10^{10}$\\
R & 55289 & 5.51$\times10^{11}$ & 3.7$\times10^{10}$\\
R & 55290 & 2.03$\times10^{11}$ &1.5$\times10^{10}$\\
R & 55291 & 5.99$\times10^{11}$ &4.3$\times10^{10}$\\
\hline
U & 55288 & 4.542$\times10^{11}$ & 6.8$\times10^{9}$\\
U & 55292 & 4.253$\times10^{11}$ & 6.3$\times10^{9}$\\
U & 55300 & 3.856$\times10^{11}$ & 6.1$\times10^{9}$\\
U & 55304 & 3.737$\times10^{11}$ & 5.5$\times10^{9}$\\
UVM2 & 55274 & 5.987$\times10^{11}$ & 8.8$\times10^{9}$\\
UVW2 & 55273 & 5.066$\times10^{11}$ & 7.9$\times10^{9}$\\
\enddata
\end{deluxetable}

\begin{deluxetable}{ccccccc}
\tabletypesize{\scriptsize}
\tablecaption{Analysis summary of the VER J0648+152 Lick Observatory Kast spectrum from 2010 November 5 (MJD 55505) \label{tbl-2}}
\tablewidth{0pt}
\tablehead{
 \colhead{Ions} &\colhead{Rest Wavelength} & \colhead{Centroid\tablenotemark{a}} & \colhead{FWHM} & Redshift\tablenotemark{b} & Observed E. W.\tablenotemark{c} & Notes \\
 \colhead{ } &\colhead{(\AA)} & \colhead{(\AA)} & \colhead{(\AA)} & Absorbed & (\AA) &  \\}
\startdata
Ca II (K) & 3934.79  & 4639.07 &20.7 &0.1789&2.60 $\pm$0.21& \\
Ca II (H)   & 3969.61& 4678.26 & 16.4&0.1785&2.47$\pm$0.19&\\
G band  & 4305.61   & 5077.46 &17.5 &0.1792&1.70$\pm$0.18&\\
Mg I        & 5174.14  &  6102.32 & 22.1&0.1793&2.35$\pm$0.20&[1]\\
Na I      & 5894.13    & 6951.66 & 23.0&0.1794&2.48$\pm$0.15&[2]\\
\enddata
\tablenotetext{a}{Based on Gaussian fit}
\tablenotetext{b}{Measured from line centroid}
\tablenotetext{c}{Error is only statistical}
\tablecomments{[1] Blanded with Mg I 5168.74 Mg I 5185.04 [2] Blanded with Na I 5891.61 and Na I 5897.57}
\end{deluxetable}

\begin{deluxetable}{cccc}
\tabletypesize{\scriptsize}
\tablecaption{SED Modeling Parameters: Summary of the parameters describing the emission-zone properties for the SSC, EC and lepto-hadronic models.  See text for parameter descriptions.\label{tbl-3}}
\tablewidth{0pt}
\tablehead{
 \colhead{Parameter} &\colhead{SSC} & \colhead{External Compton} & \colhead{Lepto-Hadronic} \\}
\startdata
$L_e$ [erg s$^{-1}$] & $7.5 \times 10^{43}$  &   $4.9 \times 10^{41}$ & $4.9 \times 10^{41}$ \\
$\gamma_1$   & $6.7 \times 10^4$  & $8.2 \times 10^4$ & $9 \times 10^3$\\
$\gamma_2$   & $10^6$            & $10^6$ &$5 \times 10^4$ \\
$q$          & $4.8$             & $4.8$  & $4.8$\\
$B$ [G]      & $0.14$           & $0.1$   & $10$\\
$\Gamma = D$ & $20$               & $20$  & $15$ \\
$ T_{EC}$ [K]  & --- & $10^3$ & --- \\
$ u_{EC}$ [erg cm$^{-3}$] & --- & $7.0 \times 10^{-8}$ & --- \\
$L_p$ [erg s$^{-1}$]           & --- & --- & $4.9 \times 10^{41}$ \\
$E_{\rm p}^{\rm min}$ [GeV]     & --- & ---   & $10^3$\\
$E_{\rm p}^{\rm max}$ [GeV]    & --- & ---   &     $1.5 \times 10^{10}$    \\
$q_p$                            & --- & ---              &  $2.0$\\
$\epsilon_{Be}$ & $0.16$ & $41$ &$1.7 \times 10^4$\\
$\epsilon_{Bp}$ & ---                     & --- & $4.2$\\
$\epsilon_{ep}$ & ---                 & ---& $2.5 \times 10^{-4}$ \\
$t_{\rm var}^{\rm min}$ [hr]  & $1.1$  & $10.9$    &      $7.2$  \\
\enddata
\end{deluxetable}

\end{document}